\documentclass[final]
  {aipproc}

\layoutstyle{6x9}

\begin{document}

\title{Global Asymmetry of the Heliosphere}


\author{Merav Opher}{
  address={George Mason University, 4400 University Drive, Fairfax, VA 22030}
}

\author{Edward C. Stone}{
  address={California Institute of Technology, Pasadena, CA 91125}
}

\author{Paulett C. Liewer}{
  address={Jet Propulsion Laboratory, California Institute of Technology, Pasadena, CA 91109}
  }
\author{Tamas Gombosi}{
  address={Center for Space Environment Modeling, University of Michigan, Ann Arbor, MI}
}

\begin{abstract}
Opher et al. \cite{opher06} showed that an interstellar magnetic field parallel to the plane defined by the deflection of interstellar hydrogen atoms can produce a north/south asymmetry in the distortion of the solar wind termination shock. This distortion is consistent with Voyager 1 and Voyager 2 observations of the direction of field-aligned streaming of the termination shock particles upstream the shock. The model also indicates that such a distortion will result in a significant north/south asymmetry in the distance to the shock and the thickness of heliosheath. The two Voyager spacecraft should reveal the nature and degree of the asymmetry in the termination shock and heliosheath.
\end{abstract}

\maketitle

\section{Introduction}

After 27 years of anticipation, Voyager 1 crossed the inward moving termination shock (TS) at $94~AU$ and is now exploring the heliosheath \cite{burlaga,decker,stone05}. The twin Voyager spacecraft are probing the northern and southern hemispheres of the heliosphere.  As Voyager 1 crossed the TS and began exploring the heliosheath it has become increasingly clear that this previously unexplored region is full of surprises. Using in situ spacecraft data to constrain the shape of the heliosphere is challenging because they are single point observations. For a quantitative global understanding of the three dimensional structure of the heliosphere, it is necessary to use modeling in conjunction with observations to probe this region.  

In mid 2002, Voyager 1 began observing strong beams of energetic termination shock particles (TSPs) streaming outward along the spiral magnetic field. The strong upstream TSP beams were observed much of the time until Voyager 1 crossed the shock at $94~AU$ in December 2004. Jokipii et al. \cite{jokipii} and Stone et al. \cite{stone05} suggested that the upstream beaming resulted from a non-spherical shock. For a spherical shock, Voyager 1 would observe upstream TSPs streaming inward along the magnetic field. With a non-spherical shock, Voyager 1 could be connected to the TS along magnetic field lines that crossed the TS (the source of TSPs) and then crossed back into the supersonic solar wind. 

In order to know Voyager 1 measurements in 2002 it is important to understand the shape of the heliosphere. The interaction of the solar system with the local interstellar medium requires an intensive modeling effort because this problem is inherently three dimensional and involves solar and interstellar magnetic fields, ionized and neutral atoms, and cosmic rays\citep{suess,zank}. The size and shape of the heliosphere will depend on the properties of both the solar wind and the local interstellar medium. 

Based on the polarization of light from nearby stars, Frisch \cite{frisch,frisch1} suggested that the magnetic field direction is parallel to the galactic plane (and directed toward $l \sim 70^{\circ}$). Voyager 3kHz radio emission data also show preferred source locations in a plane parallel to the galactic plane \cite{kur}. On the other hand, Lallement et al. \cite{lall}, mapping the solar Lyman-$\alpha$ radiation that is resonantly backscattered by interstellar hydrogen atoms, found that the neutral hydrogen flow direction differs from the helium flow direction by $4^{\circ}$. The plane of the H deflection (HDP) is tilted from the ecliptic plane by $\sim 60^{\circ}$ and is consistent with an interstellar magnetic field parallel to the HDP plane\cite{izmo}.  However, Pogorolev et al.\cite{pog06} show that the H deflection does not uniquely determine the plane of the magnetic field.

If the interstellar magnetic field is inclined to the interstellar velocity, it can produce a lateral or north-south asymmetry in the heliospheric shape \cite{pog,ratk}. Opher et al. \cite{opher06} showed that that an interstellar magnetic field parallel to the H-deflection plane (HDP) can produce a north/south asymmetry in solar wind termination shock that is consistent with Voyager 1 and Voyager 2 observations of the direction of streaming of the particles. We review these results and comment on the Heliosphere global asymmetry.

\section{Results}

The model that we used is based on the BATS-R-US code, a three-dimensional magnetohydrodynamic (MHD) parallel, adaptive grid code developed by University of Michigan\cite{gombosi} and adapted  for the outer heliosphere\cite{opher03,opher04}. The boundary conditions are described in \cite{opher06}. We considered fields parallel to the {\it HDP} with different inclination angles  $\alpha$, where $\alpha$ is the angle between the interstellar magnetic field and the interstellar wind velocity (Figure 1). 

The coordinate system in Figure 1 is the conventional one for MHD heliospheric models, with the $z$-axis as the solar rotation axis of the sun, the interstellar velocity direction in the $x$ direction, with $y$ completing the right handed coordinate system. This coordinate system is essentially the heliographic inertial coordinate system (HGI), except that in the heliographic inertial coordinates, the direction of the interstellar wind is $5.07^{\circ}$ (HGI latitude) and $178.23^{\circ}$ (HGI longitude). 
The HGI latitude for Voyager 1 and Voyager 2 become $29.1^{\circ}$  and  $-31.2^{\circ}$ relative to the interstellar wind in the $x-y$ plane, a latitude difference of $60.3^{\circ}$. Although this conventional coordinate system ignores the $5.07^{\circ}$ tilt of the solar rotation axis in respect to the interstellar wind, more significantly it ignores the tilt of the heliospheric current sheet.  
 \begin{figure}
  \includegraphics[height=.2\textheight]{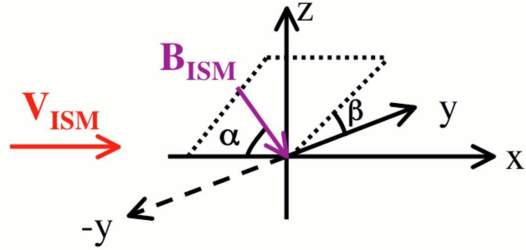}
  \caption{The coordinate system showing ${\alpha}$, the inclination angle between the interstellar wind $V_{ISM}$ and magnetic field $B_{ISM}$, and the angle ${\beta}$ between the plane of the field and the $x$-$y$ plane; ${\beta}=60^{\circ}$ for {\it HDP}. The $z$-axis is the solar rotational axis. Note that this differs from the Solar Ecliptic Coordinate (SEC) system in which the $z$-axis is perpendicular to the ecliptic plane.}
\end{figure}

Figure 2 shows the heliopause (HP) surface for $B_{ISM}$ in the {\it HDP} plane with an inclination angle $\alpha=45^{\circ}$, $B_{ISM,y}<0$, and $B_{ISM}=1.8\mu G$ (the same values are used in Figures 2-4). The heliopause 
 \begin{figure}
  \includegraphics[height=.34\textheight]{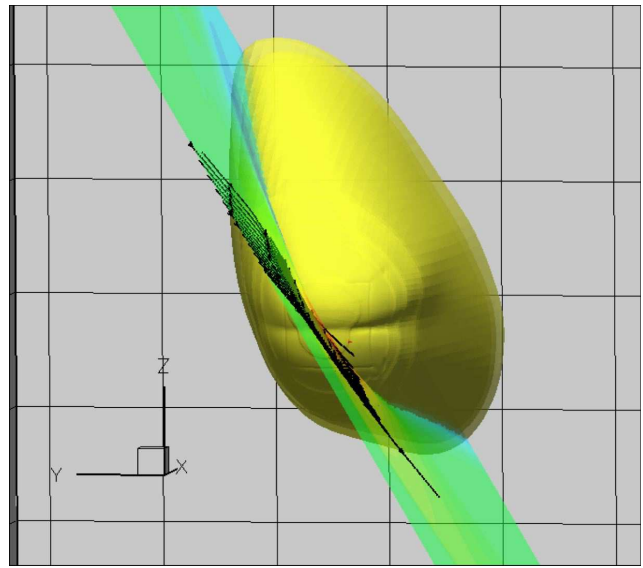}
  \caption{A model with $B_{ISM}$ in the {\it HDP} plane and ${\alpha}=45^{\circ}$ and $B_{ISM,y}<0$. The 3D iso-surface of the HP is denoted (yellow). The black lines are the interstellar magnetic field lines in the {\it HDP} plane. The contours denote magnetic field in the HDP plane }
\end{figure}
is asymmetric, north/south and east/west. It is distorted in the direction of the plane of the interstellar magnetic field and bulges outward in the northern hemisphere where the heliospheric current sheet is deflected (see also Figure 3). The current sheet deflection also results in an indentation at the solar equatorial plane that is apparent in both Figures 2 and 3. The heliosheath is thicker in the north  where the HCS is deflected. There is a smaller north/south asymmetry in the termination shock that will be discussed further below.
 
Figure 4 shows the intersection of the cones at $29.1^{\circ}$ and $-31.2^{\circ}$ that contain the spiral solar magnetic field lines crossing Voyager 1 and Voyager 2 respectively.  In both the northern and southern hemisphere, the cones intersect the surface of the termination shock nearer the equator in the nose region where the shock is closer to the Sun. The termination shock also bulges out in the northern hemisphere. As discussed in Opher et al.\cite{opher06}, $2~AU$ upstream of the shock Voyager 1 is connected to the shock along the field line in the direction toward the Sun, allowing TSPs to stream outward along the field as observed. 
 
 The distortion is larger in the southern hemisphere, so that $5~AU$ upwind of the shock Voyager 2 is connected to the shock and the TSP streaming should be inward along the spiral magnetic field line, opposite to that for Voyager 1, as has been observed \cite{cummings}.
\begin{figure}
 \includegraphics[height=.3\textheight]{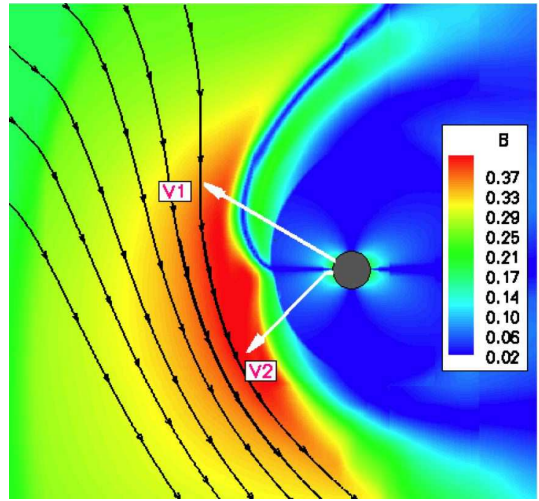}
 \caption{Contours of magnetic field strength $B(nT)$ in the  meridional plane with ${\alpha}=45^{\circ}$ and $B_{ISM,y}<0$. The black lines are the interstellar magnetic field and the white arrows denote the trajectories of Voyager 1 and Voyager 2. The heliospheric current sheet (deep blue) is deflected northward in the heliosheath. }
\end{figure}
\begin{figure}
 \includegraphics[height=.34\textheight]{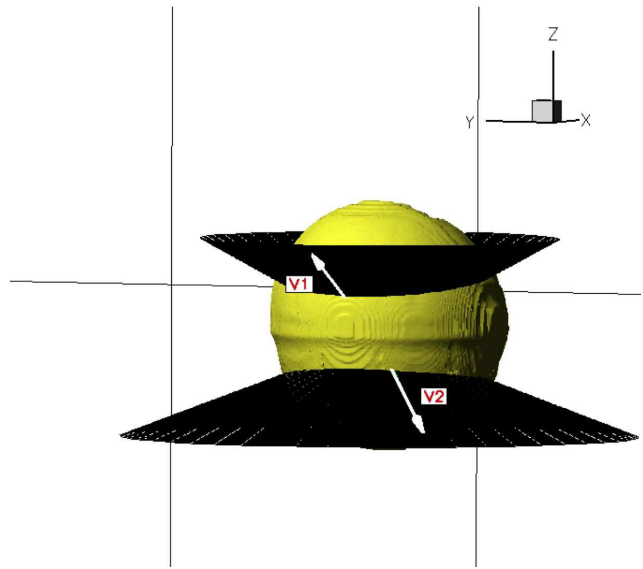}
 \caption{The intersection of the termination shock surface (yellow) with the cone of interplanetary field line connecting to Voyager 1 and Voyager 2 (black lines). The trajectories of Voyager 1 and Voyager 2 are denoted (white arrows) as well as the solar equatorial plane (red line). $B_{ISM}$ is in the {\it HDP} plane with $\alpha=45^{\circ}$ and $B_{ISM,y}<0$.}
\end{figure}
The radial distance to the shock at the latitudes of Voyager 1 and 2 decreases significantly as the inclination angle $\alpha$ increases from $\alpha =45^{\circ}$ to $60^{\circ}$, with the shock $7$ to $10~AU$ closer in the south at Voyager 2 than in the north at Voyager 1\cite{opher06}.  

Although there was no direct indication of how far Voyager 1 was from the shock when the upstream episodes of TSPs were observed, MHD models based on Voyager 2 solar wind pressure measurements \cite{rich} indicate that the distance was less than $3$ to $4~AU$. This suggests that $\alpha$ is likely in the range of $30^{\circ}$ to $60^{\circ}$ for which the distortion is maximum\cite{opher06}.
\begin{figure}
  \includegraphics[height=.3\textheight]{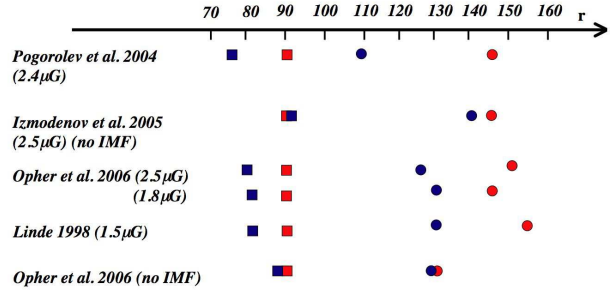}
  \caption{The red and blue squares are respectively the termination shock locations for V1 and V2 and the red and blue circles are the respective heliopause locations. The Izmodenov et al. and Opher et al. hydro models do not include an interplanetary magnetic field (IMF) and do not have a large north/south asymmetry.}
\end{figure}

Figure 5 summarizes the locations of the TS and the HP scaled from various recent models. We normalized the model results to $90~AU$ (red squares), the estimated average location of the shock at the latitude of Voyager 1 as the shock moves inward and outward over the solar cycle. The conclusions that can be drawn are: a) For the models including an interplanetary magnetic field, the TS ranged between $76$ to $82~AU$ for Voyager 2; b) There is somewhat less North-South asymmetry with a weaker $B_{ISM}$; c) The heliosheath thickness at Voyager 1 is about $55$ to $59~AU$. For the model that has an interplanetary magnetic field and strong $B_{ISM}$, the thickness at Voyager 2 is $33$ to $45~AU$. Although this is a preliminary calculation and does not incorporate a tilted current sheet, it does indicate the possibility of a significant north-south asymmetry in the heliosheath thickness; d) These models do not include neutrals that may affect the degree of asymmetry. However, models such as Linde 1998 \cite{linde98} (same model as in Linde et al. 1998 \cite{linde}) that include neutrals as a fluid show a north-south asymmetry with a ratio of $\sim 1.13$ for the north/south distance to the termination shock and $1.44$ for the ratio for the HP distance, similar to models without neutrals. Izmodenov et al. \cite{izmo}, using a kinetic treatment with the $B_{ISM}$ in the {\it HDP} plane, found a smaller north-south asymmetry. Their treatment, however, did not include the solar magnetic field. 

\section{Conclusions}
The recent results by Opher et al.\cite{opher06} indicate that an interstellar magnetic field 
$(B\sim 2\mu G)$ in the H-deflection plane can distort the heliosphere in a manner consistent with the streaming observed by Voyager 1 and Voyager 2. The model also indicates that such a distortion will result in a significant north/south asymmetry in the distance to the shock and the thickness of heliosheath, and it is reasonable to expect that Voyager 2 will encounter the shock in the next two years. When combined with a more realistic heliospheric model with a tilted current sheet, this should provide additional information about the strength and inclination of the local interstellar magnetic field. 

\begin{theacknowledgments}
The authors would like to thanks the use of Columbia cluster NASA Ames. Part of this work is the result of research performed at the Jet Propulsion Laboratory of the California Institute of Technology under a contract with the National Aeronautics and Space Administration.
\end{theacknowledgments}

\bibliographystyle{aipprocl} 

\IfFileExists{\jobname.bbl}{}
 {\typeout{}
  \typeout{******************************************}
  \typeout{** Please run "bibtex \jobname" to optain}
  \typeout{** the bibliography and then re-run LaTeX}
  \typeout{** twice to fix the references!}
  \typeout{******************************************}
  \typeout{}
 }

\end{document}